\documentclass[aps,prb,reprint,superscriptaddress]{revtex4-2}
\usepackage{graphicx} 
\usepackage{dcolumn} 
\usepackage{bm} 
\usepackage[mathlines]{lineno}

\usepackage{amsmath}
\usepackage{amssymb}
\usepackage{xcolor}

\begin{document}
	\title{Microscopic study of supercurrent diode effect in chiral nanotubes}
	\author{Chuang Li}
	\affiliation{International Center for Quantum Design of Functional Materials (ICQD), 
		Hefei National Research Center for Interdisciplinary Sciences at the Microscale, 
		University of Science and Technology of China, Hefei, Anhui 230026, China 
	}
	\author{James Jun He}
	\affiliation{Hefei National Laboratory, Hefei, Anhui 230088, China	}
	\date{\today}

\begin{abstract}
	Nonreciprocity of supercurrents may exist when both spatial inversion and time-reversal symmetries are broken, leading to the supercurrent diode effect (SDE). The spatial inversion symmetry may be broken by chiral structures in nanotubes where the SDE is expected when a magnetic flux passes through the tube. While such an effect has been predicted based on a phenomenological theory, a microscopic and quantitative study with a concrete lattice model is missing. Here, we investigate the SDE in chiral nanotubes made of carbon and those made of transition metal dichalcogenides (TMD) with tight-binding models. We obtain the  SDE efficiency as a function of the nanotube radius, the chiral angle,  the magnetic flux, the temperature, the chemical potential, etc., and find that sign flipping happens in various parameter dependencies. In TMD nanotubes, the SDEs with and without the spin-orbit coupling are compared.  We also simulate CNTs made from square lattice materials for comparison and discuss the effects of strains. Besides qualitative consistency with previous phenomenological theory, new features are found and the microscopic origins are clarified.  
\end{abstract}
\maketitle
	
	
	\section{Introduction}
	
Noncentrosymmetric quantum materials \cite{Tokura-2018-NatCom} may exhibit nonreciprocal perperties where intrinsic physical observables vary as the corresponding external perturbations are inverted. Such observables include electrical or spin conductivity, optical absorption, refractive index, critical supercurrent, etc. Particularly, nonreciprocity in superconducting systems \cite{RWakatsuki-2017-SciAdv,Tokura-2018-NatCom,RWakatsuki-2018-PRL,SHoshino-2018-PRB,YMItahashi-2020-SciAdv}, especially the supercurrent diode effect (SDE) \cite{Ando-2020-Nat,YFanQi-2022-PNAS,ADaido-2022-PRL,HJun-2022-NJP,Nadeem-2023-NatRevPhys}, has drawn intense research interest for its potential application in nondissipative and quantum coherent devices. 

The SDE refers to the phenomenon where the critical currents in opposite directions have different magnitudes, {\it i.e.} $I_{c+} \neq I_{c-}$. Obviously, the spatial inversion symmetry must be broken to observe this phenomenon. And, since the supercurrent originates from a coherent quantum state without nonequilibrium processes, the time-reversal symmetry must also be broken to achieve the SDE. Experimentally, this effect is often observed in superconducting systems under external magnetic field \cite{Masuko-2022-npjQM,Ando-2020-Nat,YMItahashi-2020-SciAdv,STimo-2020-PRB,LYangYang-2021-NatCom,Miyasaka-2021-APE,EStrambini-2022-NatCom,Bauriedl-2022-NatCom,BPal-2022-NatPhys,BChristian-2022-NatNano,Kawarazaki-2022-APE,GMohit-2023-NatCom,YHou-2023-PRL,JDiezMerida-2023-NatCom,Gutfreund-2023-NatCom,YJonginn-2023-PRR,Kealhofer-2023-PRB,Anh-2024-NatCom} or with internal magnetic order \cite{Narita-2022-NatNano,KRJeon-2022-NatMat,Trahms-2023-Nat,Narita-2023-SciAdv} (with a few exceptions \cite{LJXiaZi-2022-NatPhys,WHeng-2022-Nat,Golod-2022-NatCom,LTian-2024-Nat} which are still to be understood). On the other hand, the inversion-symmetry breaking may come from noncentrosymmetric crystal structures or interface effect which leads to spin-orbit coupling (SOC) \cite{HJun-2022-NJP,ADaido-2022-PRL,ADaido-2022-PRB,YFanQi-2022-PNAS,SIliic-2022-PRL,dePicoli-2023-PRB}. It may also happen that both symmetries are broken together by the same mechanism, such as the valley polarization \cite{ZhBaoXing-2022-PRB,HJinXin-2023-PRL,XYingMing-2023-PRR}. 

Chiral atomic structrures can break the inversion symmetry in a different way without introducing SOC. It remained unclear weather the SDE exists in such systems until a recently phenomenological study gave an affirmative answer for chiral nanotubes (CNTs) with magnetic fluxes \cite{HJun-2023-NatCom}. The SDE in CNTs were found to depend on the chiral angle, tube radius, the magnetic flux, etc. However, the discussion was limited to CNTs formed by square-lattice materials whereas realistic nanotubes usually appear as rolled triangular/honeycomb lattices, such as carbon and transition metal dichalcogenide (TMD) nanotubes \cite{FQin-2017-NatCom}. A microscopic and quantitative investigation on the SDE in CNTs is needed. 

Here, we study the SDE in single-walled chiral nanotubes made of carbon and transition metal dichalcogenides with tight-binding models. We show how the SDE efficiency varies with parameters such as nanotube radius, chiral angle, magnetic flux, temperature, and chemical potential. Particularly, it is found that the SDE can flip signs depending on these parameters and the effect is influenced by the presence of spin-orbit coupling in TMD nanotubes. The research provides a microscopic understanding of the SDE, showing qualitative consistency with previous phenomenological theories while uncovering new features specific to chiral nanotubes.

The remaining of this manuscript is organized as follows. In Section II, we describe the geometric structure of chiral nanotubes, using carbon nanotubes as an example, and explain how critical currents are calculated. The impact of a magnetic field on the dispersion of chiral nanotubes is analyzed, showing how it leads to asymmetric dispersion and the breaking of inversion symmetry.
In Section III, we focus on superconducting chiral carbon nanotubes. We investigate the variation of the SDE coefficient ($\eta \equiv (I_{c+}-I_{c-})/(I_{c+}+I_{c-})$) with nanotube geometry, including circumferential vectors, radius, and chiral angle. 
In Section IV, we extend our study to chiral TMD nanotubes, say MoSe$_2$. The variation of $\eta$ with temperature and magnetic flux is analyzed, both in the presence and absence of spin-orbit coupling (SOC). The impact of chemical potential on $\eta$ is examined.
In Section V, as a comparison, we investigate the SDE in chiral nanotubes composed of square lattice materials. Differences in the periodicity and smoothness of $\eta$ compared to triangular/honeycomb lattices are noted. Section VI is a discussion of possible effects of the strains. 
Finally, we summarize our findings and discuss the relevance to real materials in Section VII.

	
	\section{Models and methods}\label{sec:CT}
	
\begin{figure}
	\centering
	\includegraphics[width=8.5cm]{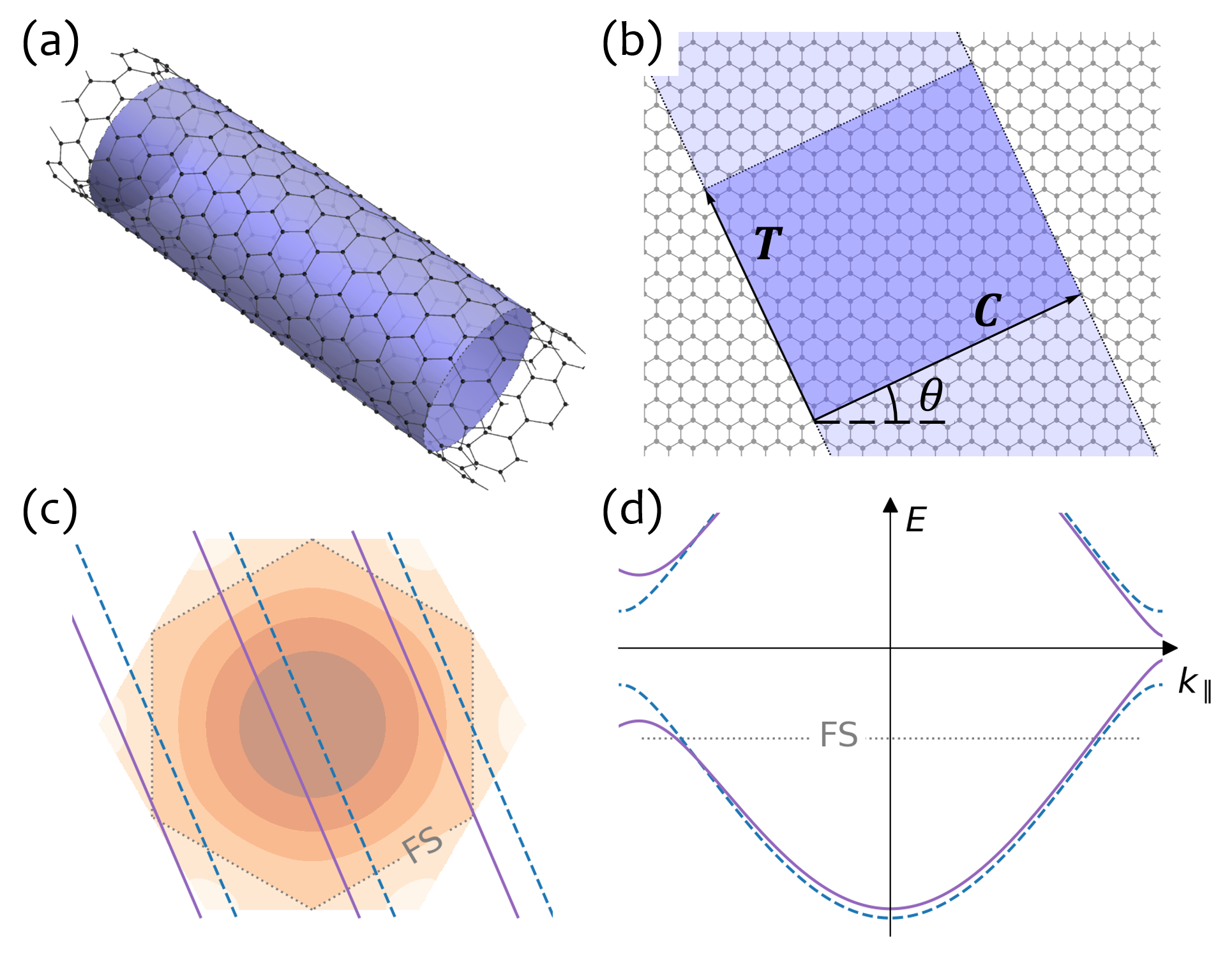}
	\caption{\label{fig:Tube-Illus}(a) Illustration of chiral carbon nanotubes. (b) The Chiral carbon nanotube is unfolded into a two-dimensional plane, becoming a ribbon on the graphene. The deep blue rectangle forms a super unit cell, corresponding to the blue region in (a). (c) Dispersion of graphene in the first Brillouin zone, and the corresponding momentum $\mathbf{k}$ of chiral carbon nanotubes in the absence (dashed line) and presence (solid line) of magnetic flux through the tube. (d) Asymmetric (symmetric) partial dispersion of chiral carbon nanotubes in the longitudinal direction in the presence (absence) of a magnetic field.}
\end{figure}
	
	Chiral nanotubes are microscopic tubular structures in which the lattice basis vectors are not parallel to the circumference. 
	To understand the lattice structure of chiral nanotubes, taking the carbon nanotube as an example, we first cut the tube wall along the cylindrical axis, and unfold it into a rectangular ribbon on a honeycomb lattice in two dimensions (2D), as shown in Fig.~\ref{fig:Tube-Illus}(a) and (b). The curvature effect is ignored here.
    On the unfolded two-dimensional lattice,  the primitive vectors are $\hat{\mathbf{a}}_1=(1,0)$, $\hat{\mathbf{a}}_2=(1/2,\sqrt{3}/2)$. The lattice constant is taken as unity. 
    The transverse vector $\mathbf{C}$ along the circumferential direction and the longitudinal vector $\mathbf{T}$ along the axial direction form the basis vectors of the super unit cell. 
    A CNT can be uniquely determined by its circumferential vector $\mathbf{C}=m\hat{\mathbf{a}}_1+n\hat{\mathbf{a}}_2$ and let us label it by the geometric indexes $(m,n)$.
    The angle between $\mathbf{C}$ and the unit vector $\bm a_1$  is the chiral angle $\theta$.
    Once $\mathbf{C}$ is determined, the geometry of the nanotube, i.e., $\theta$ and the radius $R$ are determined through $\tan \theta = \sqrt{3}n/(2m+n)$ and $R=|\mathbf{C}|/2\pi=\sqrt{m^2+n^2+mn}/2\pi$. The longitudinal unit vector $\mathbf{T}$ can also be obtained from $\mathbf{C}\cdot\mathbf{T}=0$. 

    In momentum space, the finite circumferential vector makes the wave vector $\mathbf{k_\perp}$ in that direction discrete compared to graphene, and thus chiral nanotubes correspond to quasi-1D structures therein whose $\mathbf{k_\perp}$ satisfies $\exp(i\mathbf{k_\perp} \cdot \mathbf{C})=1$, as shown in Fig.~\ref{fig:Tube-Illus}(c). 
    In the absence of an external field, its dispersion is symmetric along the axial direction, as shown by the dashed line in Fig.~\ref{fig:Tube-Illus}(d). However, when there is a magnetic flux passing through the chiral nanotube, i.e. $\Phi=\int \bm A \cdot d\bm l$, the magnetic vector potential $\mathbf{A}$ along the circumferential direction will shift $\mathbf{k_\perp}$ as shown by the solid line in Fig.~\ref{fig:Tube-Illus}(c), and the dispersion becomes asymmetric [Fig.~\ref{fig:Tube-Illus}(d)].

    The key to one of the main methods of generating the superconducting diode effect is to break all symmetries between the two opposite directions of the system before applying a current. For superconducting chiral nanotubes, the chiral angle has broken the 3D inversion symmetry $(R,\varphi,z)\rightleftharpoons(R,\varphi+\pi,-z)$, where the cylindrical coordinates are used with $z$ along the axial direction and $\varphi= 2\pi \mathbf{r}\cdot\mathbf{C}/|\mathbf{C}|^2$ being the azimuth angle. Further, by introducing magnetic flux through the tube, the magnetic field will break the time-reversal symmetry, and meanwhile, the chiral angle and the magnetic vector potential  together break the pseudo-inversion symmetry $(R,\varphi,z)\rightleftharpoons(R,-\varphi,-z)$, which is the 2D inversion symmetry operating on the unfolded nanorubon, as shown by the solid lines in Fig.~\ref{fig:Tube-Illus}(c) and (d). As all symmetries connecting $z$ and $-z$ are broken, we expect the SDE along the tube. 
	
To calculate the critical supercurrents, we consider a tight binding model for an infinitely long superconducting CNT which has a super unit cell determined by $(m,n)$. 
Using the basis $\bar{c}_{i\nu}^\dagger= ( c_{i\nu\uparrow}^\dagger,\, c_{i\nu\downarrow} )$, where $i$ labels the sites on the Bravais lattice in each supercell, $\nu $ denotes the sublattice (for graphene) or orbital (for TMD) degrees of freedom and $\uparrow,\, \downarrow$ denotes the spin, the tight-binding Hamiltonian can be written as
\begin{align}\label{eq:Hmtn-graphene}
	H =& \sum_{ \langle ij \rangle } \bar{c}_{i\nu}^\dagger  \hat{t}_{ij}^{\nu \nu'}  \bar{c}_{j\nu'} + \text{H.c.}
	+ \sum_{i\nu} \bar{c}_{i\nu}^\dagger \begin{pmatrix}
		-\mu & \Delta_\nu \\ \Delta_\nu^* & \mu
	\end{pmatrix} \bar{c}_{i\nu} \ ,
\end{align} 
where $\mu$ is the chemical potential, $\Delta$ is the superconductivity order parameter, and the hopping matrix
\begin{align}\label{eq:t}
	\hat{t}_{ij}^{\nu \nu'} = t_{\nu \nu'}(\mathbf{d}) \begin{pmatrix}
		-e^{-i\left( \mathbf{K} +\mathbf{q}/2 +\tilde{\mathbf{A}} \right) \cdot  \mathbf{d}} & 0 \\
		0 &  e^{i\left( -\mathbf{K} +\mathbf{q}/2 +\tilde{\mathbf{A}} \right) \cdot  \mathbf{d}}
	\end{pmatrix} \ .
\end{align}
The subscript ${ \langle ij \rangle }$ means summation over neighboring sites and $\mathbf{d}$ is the displacement of the hopping (depends on $i,j$).  
The wave vector $\mathbf{K}$ is the quantum number that corresponds to the periodicity of the chiral nanotube supercell along the axial direction. 
The Cooper pair momentum $\mathbf{q}$ along the axial direction may be nonzero when a current flows. 
We defined $\tilde{\mathbf{A}}=e\mathbf{A}/\hbar$  with $\mathbf{A}$ being the vector potential along the circumference.  In presence of a flux $\Phi$ through the nanotube, we have $A = \Phi/(2\pi R)$. 

For carbon CNTs,  there are three nearest neighbors connected by $\mathbf{d}_1=\mathbf{0}, \mathbf{d}_2=\hat{\mathbf{a}}_1-\hat{\mathbf{a}}_2$ and $\mathbf{d}_3=-\hat{\mathbf{a}}_1-\hat{\mathbf{a}}_2$. And the hopping
\begin{align} \label{eq:tC}
	t_{\nu \nu'}^\text{C} = -t \begin{pmatrix}
		0 & 0 \\
		1 & 0
	\end{pmatrix}_{\nu\nu'} \ . 
\end{align}
For TMD CNTs, the hopping terms are more complicated and given in Appendix \ref{Appendix1}. 
	
	The  order parameter may also vary along the transverse direction, which leads to a phase dependent on the azimuth angle, i.e. 
	\begin{align}
		\Delta_\nu = \Delta_0 \exp(in_S\varphi) \ ,
	\end{align}
	where $n_S \in \mathbb{Z}$. Similar to the Little-Parks effect \cite{LittleParks,Tinkham-2004-book}, $n_S$ is non-zero when a sufficiently large magnetic flux is passed through the nanotube, and rises stepwise as the flux increases.
	
	\begin{figure}[tb]
		\centering
		\includegraphics[width=8.5cm]{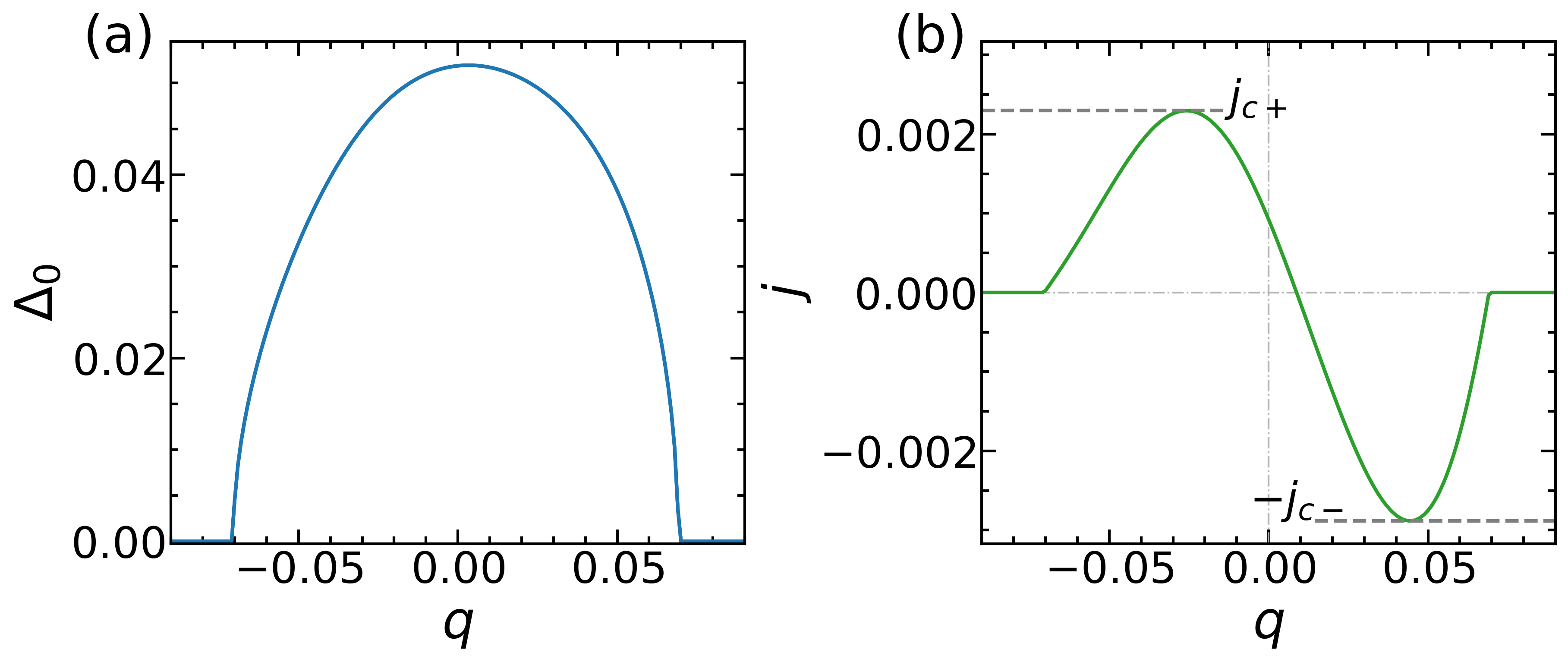}
		\caption{\label{fig:Delta-js-q}Variation of (a) the superconducting gap $\Delta_0$ and (b) the supercurrent with Cooper pair momentum $q$ in chiral carbon nanotubes. The fixed parameters of the nanotubes are $(m,n)=(12,6)$, $\mu=-2.7$, $\tilde{\Delta}_0=0.1$, $\Phi=0.3\Phi_0$, $T=0.03$.}
	\end{figure}

    We calculate the SDE in the nanotube system as follows (see Appendix \ref{Appendix2}).
    First, we  determine the superconducting gap $\Delta_0$ at certain tube-directed Cooper pair momentum $q$. This can be done by solving the self-consistency equation 
    \begin{align} \label{eq:Gap-Eq}
        \Delta_\nu = -U \langle c_{\mathbf{r} \nu \downarrow} c_{\mathbf{r} \nu \uparrow} \rangle \ .
    \end{align}
    Anderson-acceleration is helpful in this fix-point calculation \cite{NPeng-2011-SIAM, TAlex-2015-SIAM}. 
    $U>0$ is the attraction potential between the electrons. For the given nanotube geometry, we assume that $U$ do not depend on $q$, $\Phi$ and $T$, the temperature.
	For a given the zero-temperature zero-field superconducting gap $\tilde{\Delta}_{0}$, the corresponding $U$ can be calculated directly using Eq.~\ref{eq:Gap-Eq}. 
	
	Having determined the superconducting gap $\Delta_0$, and consequently the Hamiltonian Eq.~\eqref{eq:Hmtn-graphene}, the current density along the superconducting nanotube is given by
	\begin{align}
		j(q)
		=-\frac{2e}{\hbar S} \int_{0}^{2\pi} \frac{dK}{2\pi} 
		\sum_\alpha \langle \psi_\alpha | \frac{\partial H}{\partial q} | \psi_\alpha \rangle f_D(E_\alpha) \ ,
	\end{align}
    where 
    $S=|\mathbf{C}||\mathbf{T}|$ is the area of the supercell. $f_D(E_\alpha)$ is the Dirac-Fermi distribution of the eigen state $| \psi_\alpha \rangle$ with the eigen energy $E_\alpha$.
	As shown in Fig.~\ref{fig:Delta-js-q}(b), $j$ varies with $q$, where the maximum and minimum values are the critical current densities $j_{c+}$ and $-j_{c-}$ along opposite directions.
    The SDE strength is given by 
	$
		\eta = (j_{c+}-j_{c-})/(j_{c+}+j_{c-}).
	$
	When the magnetic flux $\Phi/\Phi_0$ is in the vicinity of a half-integer, it is necessary to calculate the supercurrent that can be reached in the system for all possible $n_S$, to determine the critical currents of the system. 
	
	Figure~\ref{fig:Delta-js-q} shows the complete curves of the superconducting gap $\Delta_0$ and supercurrent $j$ as a function of the Cooper pair momentum $q$. $U$ is fixed during the variation.
	When $q$ is small, $j$ is approximately proportional to $q$; when $q$ is large, $\Delta_0$ decays rapidly and $j$ drops to zero due to the breaking of Cooper pairs.
	Under competition between the increasing velocity and the decreasing number of Cooper pairs, $j$ reach the critical currents $j_c$ somewhere.
	For the practical calculation of the SDE coefficients, we seek for the current extrema by optimization.
	Due to the asymmetry, $j_{c\pm}$ may not be equal. And also, the ground state, where the free energy is minimized as $j\propto -\partial F/\partial q =0$, has an unvanishing $q$, as shown in Fig.~\ref{fig:Delta-js-q}(b).
	
	\section{SDE in Carbon CNT}

	We first concentrate on superconducting  carbon CNTs.
	Referring to graphene, we use hopping strength $t=2.7$, in unit of eV \cite{ReichS-2002-PRB}. In order to explicitly show the numerical results, we first set $\mu=-2.7$ and $\Delta=0.1$, and will discuss the influence of these parameters later.
	
\begin{figure}[tbh]
		\centering
		\includegraphics[width=8.5cm]{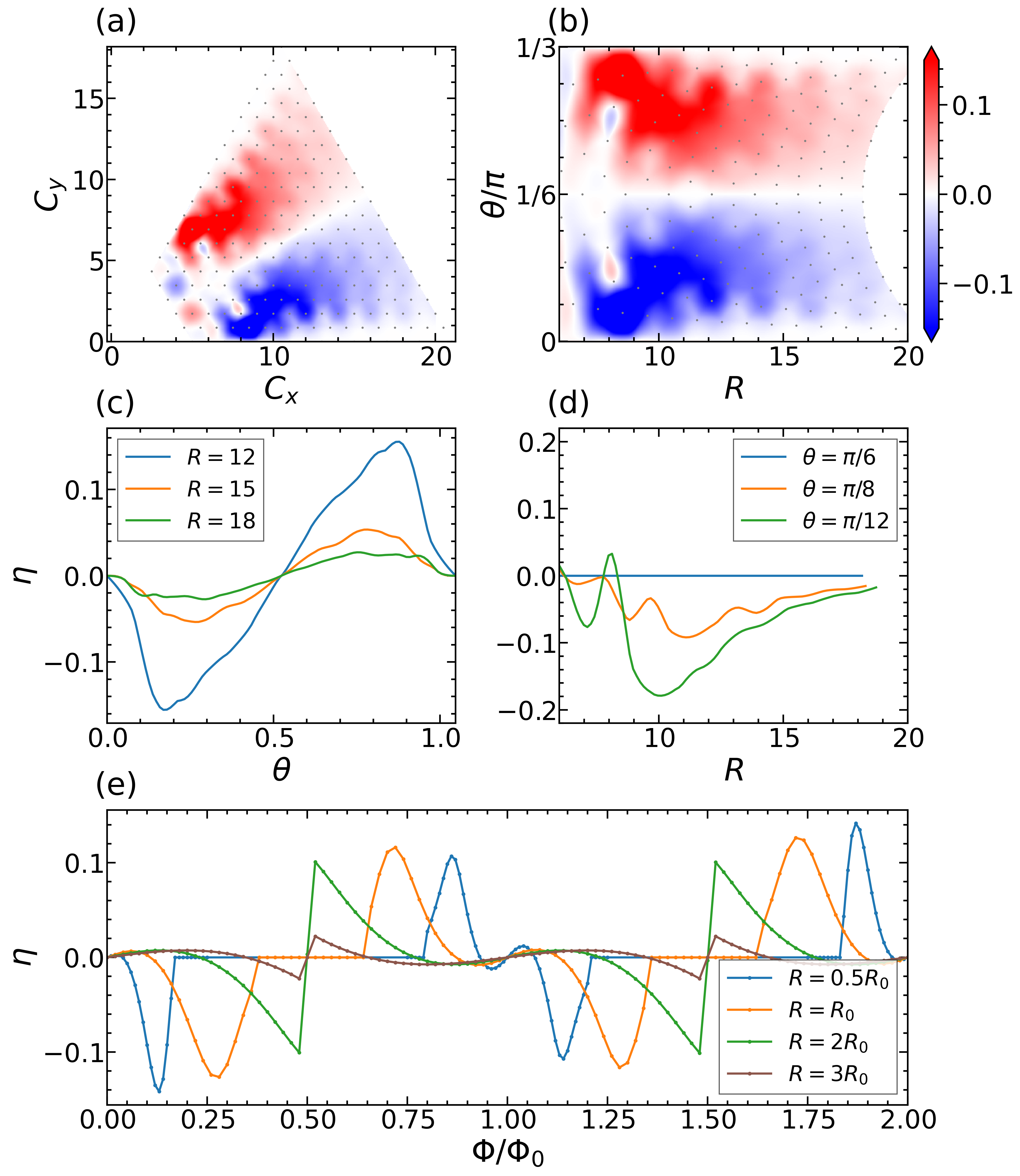}
		\caption{\label{fig:HC-SDE-geo}Geometry-dependent SDE in chiral carbon nanotubes.  (a) Variation of SDE coefficients $\eta$ with circumferential vector $\mathbf{C}$. (b) Variation of $\eta$ with nanotube radius $R$ and chiral angle $\theta$. The dots in (a) and (b) denotes the data points. (c) Variation of $\eta$ with $\theta$ at different radii. (d) Variation of $\eta$ with $R$ at different chiral angles. (e) Comparison of $\eta(\Phi)$ at different radii, where $R_0$ corresponds to geometric index $(12,6)$. The parameters are $t=2.7$, $\mu=-2.7$, $\tilde{\Delta}_0=0.1$, $\Phi=0.18\Phi_0$ and $T=0.03$, unless specified otherwise.}
	\end{figure}

	To investigate the variation of SDE with nanotube geometry, we calculated the coefficient $\eta$ for different circumferential vectors $\mathbf{C}$, resulting in Fig.~\ref{fig:HC-SDE-geo}(a). For each geometric structure, we varied the interaction $U$ to keep $\Delta$ constant. Each gray point corresponds to a structure of nanotubes, and the colors are obtained by interpolation. 
	Converting the argument $\mathbf{C}$ to the corresponding radii $R$ and chiral angles $\theta$, the variation of SDE with respect to the latter two is shown in Fig.~\ref{fig:HC-SDE-geo}(b$\sim$d). 
	Since the dispersion of graphene exhibits $C_6$ symmetry [Fig.~\ref{fig:Tube-Illus}(c)], $\eta(\theta)$ will also have a period of $\pi/3$.
	Within the period, $\eta(\theta)$ is similar to a sinusoidal curve. At special angles $\theta=0$, $\pi/6$, and $\pi/3$, the inversion symmetry in the nanotube is restored and the SDE vanishes.
	Nanotubes with opposite chirality have opposite SDE, i.e., $\eta(\theta)=-\eta(-\theta)=-\eta(\pi/3-\theta)$.
	Fixing $\theta$, as the radius $R$ increases, the absolute value of the SDE coefficient $|\eta|$ increases and then decreases, with strong oscillation due to the finite size effect.
	
	To better understand the variation, we need to observe the variation of SDE with flux $\Phi$ in the tube for different $R$, as shown in Fig.~\ref{fig:HC-SDE-geo}(e).
	In the interval of small magnetic field $\Phi<\Phi_0/2$, for sufficiently small nanotubes, as $\Phi$ increases, $\eta$ first increases slightly, then changes sign and produces a larger extreme value, and finally disappears when the critical magnetic field is reached. 
	The smaller the nanotube, the smaller the critical flux that the superconductor can tolerate, which is simply due to the fact that a smaller tube with fixed flux means a larger magnetic field. The entire pattern of the $\eta(\Phi)$ curve shrink horizontally as $R$ is changed. On the contrary, as the nanotube size increases, $\eta(\Phi)$ stretches and gets truncated at $\Phi_0/2$ before reaching the negative  extremum.
	The $\eta(\Phi)$ curve pattern described here is common in triangular Bravais lattices, as we shall see below.

	\begin{figure}
	\centering
	\includegraphics[width=8.5cm]{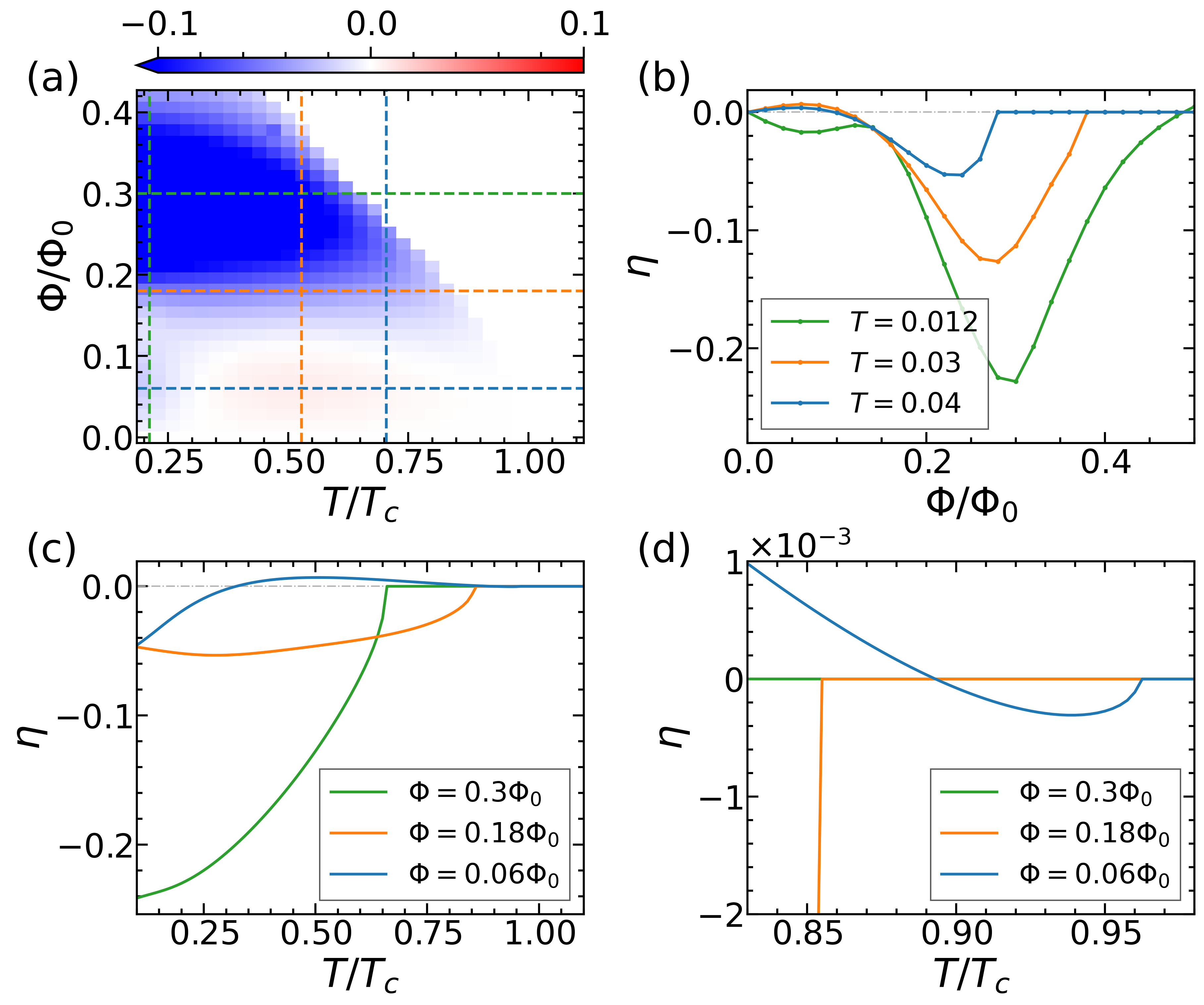}
	\caption{\label{fig:HC-SDE-temp-flux}(a) Variation of the SDE coefficient $\eta$ with temperature $T$ and magnetic flux $\Phi$ in the chiral carbon nanotubes with geometric index $(12,6)$. (b)/(c) Variation of $\eta$ with $\Phi$/$T$ for different temperatures/fluxes, corresponding to the dashed lines in (a). (d) is a zoom in of (c) near $T_c$. Other fixed parameters are the same as in Fig.~\ref{fig:HC-SDE-geo}.}
\end{figure}

	Considering a wider range of magnetic fields, $\eta(\Phi)$ is strictly in $2\Phi_0$ period, since the Hamiltonian matrix $H(\Phi, n_S=0)$ is similar to $H(\Phi+2\Phi_0, n_S=-2)$, just like the flux periodicity in Little-Parks effect. 
	Since $\eta$ changes sign when the magnetic field is flipped, one has $\eta(\Phi)=-\eta(-\Phi)=-\eta(2\Phi_0-\Phi)\approx -\eta(\Phi_0-\Phi)$,  and thus $\eta$ flips its sign at $\Phi_0/2$ if the critical field is not reached, which is possible if  $R$ is large.
	
\begin{figure}
	\centering
	\includegraphics[width=8.5cm]{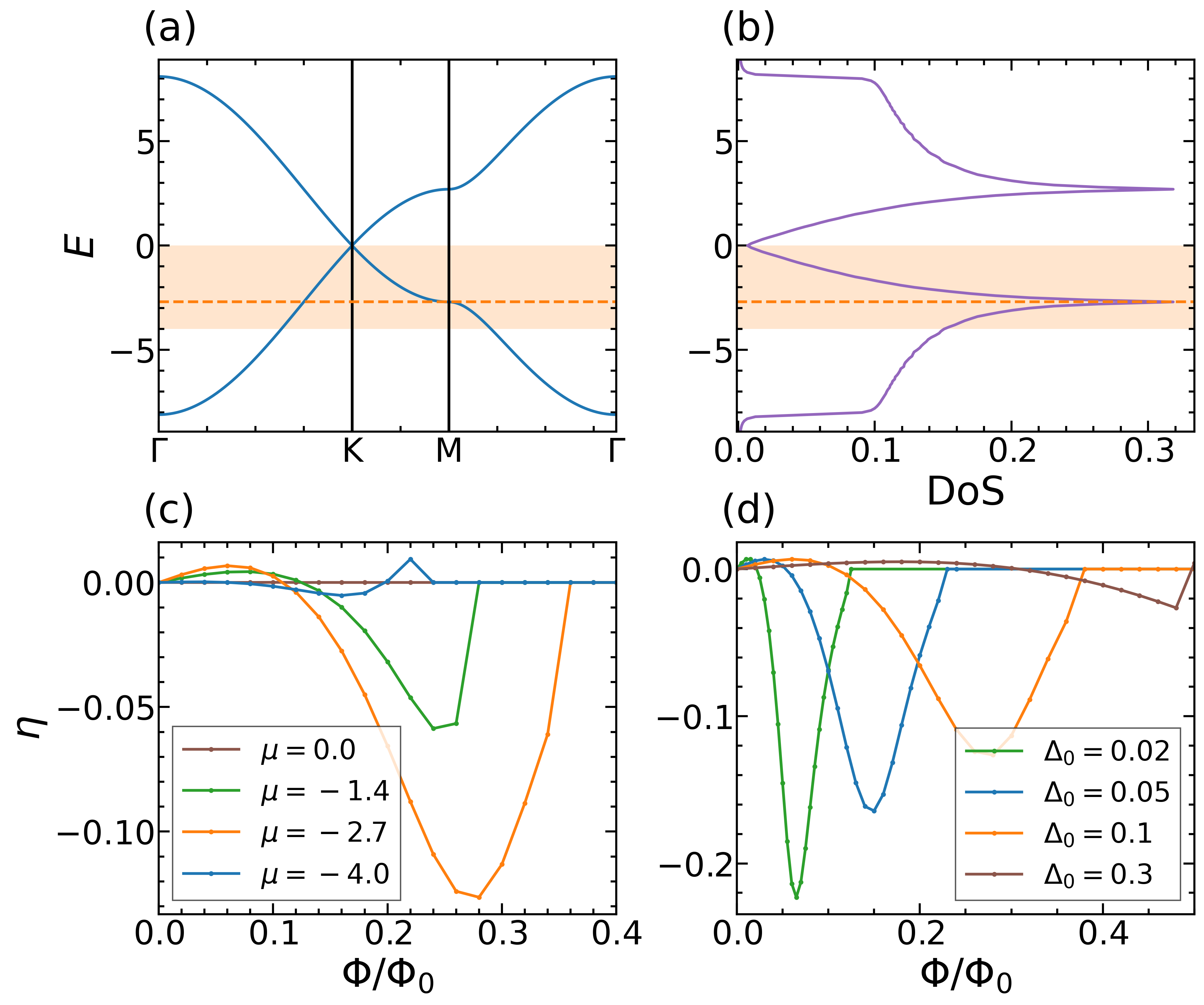}
	\caption{\label{fig:HC-SDE-mu-delta0}(a) Dispersion and (b) density of states (DoS) of graphene, with the dashed line showing the position of the Fermi surface in the previous calculation. (c) SDE $\eta(\Phi)$ curves at different Fermi surfaces corresponding to the orange regions in (a) and (b). (d) SDE $\eta(\Phi)$ at $\mu=-2.7$ for different $\tilde{\Delta}_0$, temperature also changes as $T=0.3\tilde{\Delta}_0=0.528T_c$. Other fixed parameters are the same as in Fig.~\ref{fig:HC-SDE-temp-flux}.}
\end{figure}
    Figure~\ref{fig:HC-SDE-temp-flux} shows the variation of the SDE with temperature $T$ and magnetic flux $\Phi$ for a $(12,6)$  carbon CNT  ($R\approx 16$ and $\theta \approx 0.1\pi$). 
    The critical temperature at zero field is estimated as $T_c=\tilde{\Delta}_0/1.76$.
    In a certain range below the critical temperature where the magnetic field is relatively weak, $\eta$ is positive; while in the region with relatively strong magnetic field and low temperature, $\eta$ has the opposite sign and can take a larger magnitude.
    As shown in Fig.~\ref{fig:HC-SDE-temp-flux}(c) and the partial enlargement (d), whatever the field is, near the critical temperature, $\eta$ shows a square-root-like temperature dependence, as described by the Ginzburg--Landau theory \cite{HJun-2022-NJP}.
	
\begin{figure*}[tbh]
		\centering
		\includegraphics[width=16cm]{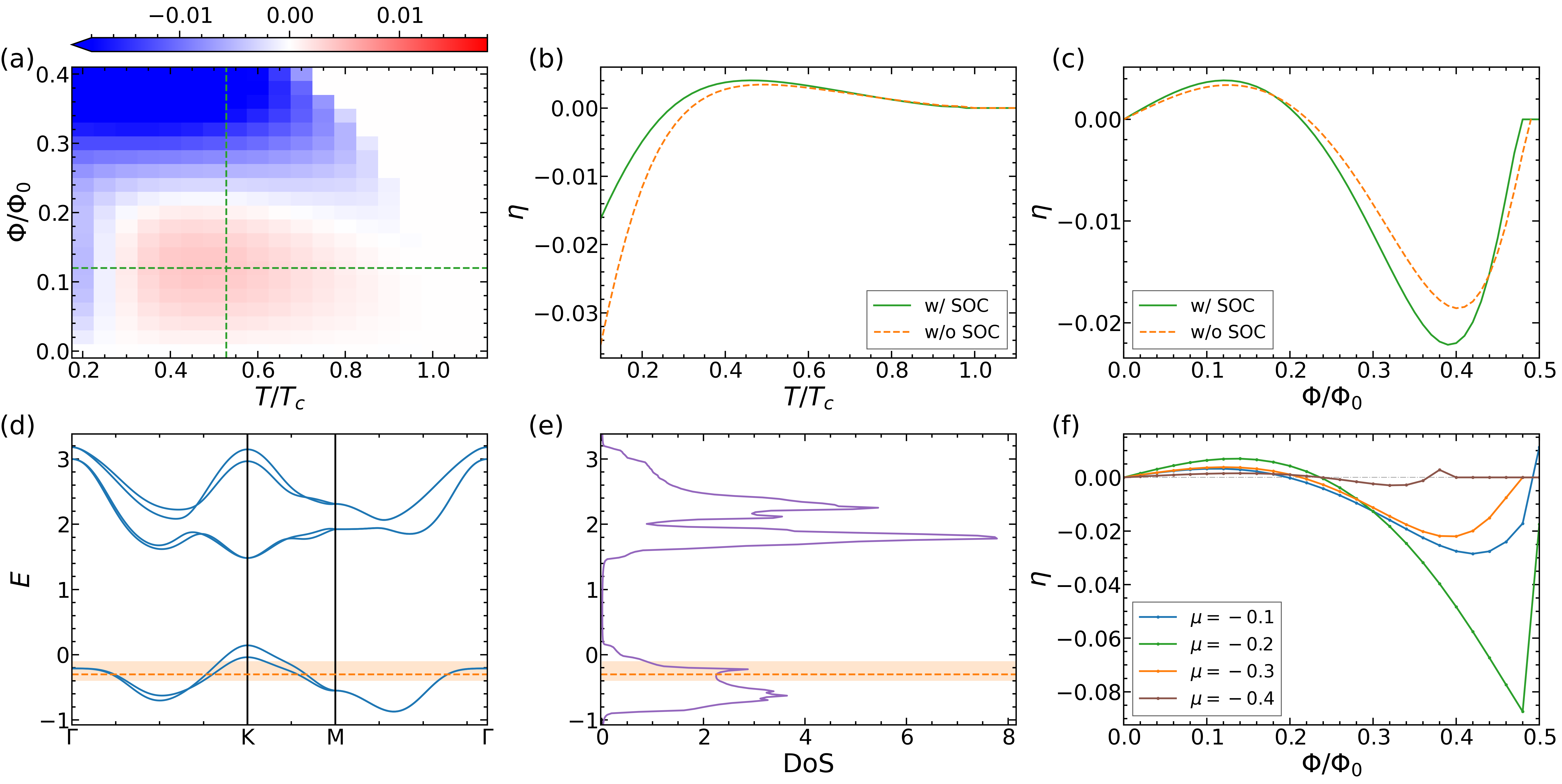}
		\caption{\label{fig:TMD-SDE}SDE in chiral TMD (MoSe$_2$) nanotubes. (a) Variation of the SDE coefficient $\eta$ with temperature $T$ and magnetic flux $\Phi$. Variation of $\eta$ with (b) $T$ and (c) $\Phi$ in the presence and absence of SOC. (d) Dispersion and (e) density of states of graphene, with the dashed line showing the position of the Fermi surface in the previous calculation. (f) $\eta(\Phi)$ curves at different Fermi surfaces corresponding to the orange regions in (d) and (e). The parameters of MoSe$_2$ are taken from the three-band model with up to third-nearest-neighbor hoppings in generalized-gradient approximation case \cite{LGuiBin-2013-PRB}. Other parameters are geometric index $(24,12)$, $\mu=-0.3$, $\tilde{\Delta}_0=0.02$, $\Phi=0.12\Phi_0$, and $T=0.006$, unless specified.}
\end{figure*}
	
	The SDE effect is also strongly affected by the chemical potential $\mu$. 
	Fig.~\ref{fig:HC-SDE-mu-delta0}(a) and (b) show the dispersion and density of states of unfolded graphene. The chemical potential $\mu$ of the nanotube system in previous calculations was set near the van Hove singularity as shown by the dashed line. The $\eta(\Phi)$ curves for several values of $\mu$ inside the orange region are shown in Fig.~\ref{fig:HC-SDE-mu-delta0}(c). One sees that $\eta$ varies drastically with $\mu$. It is tiny when $\mu$ is near the Dirac point, and the largest SDE, $|\eta|\approx 13\%$, is reached at the van Hove singularity. As $\mu $ approaches the van Hove singularity, the Fermi surface closer and closer to a perfect hexagon, which is the most anisotropic case. Such an anisotropy combined with the chiral structure is the key to the nonreciprocity here. 
	
	For the sake of practicality we also consider the effect of the superconducting gap $\tilde{\Delta}_0$ on $\eta$. 
	As $\tilde{\Delta}_0$ decreases, the $\eta(\Phi)$ curve shrinks horizontally as shown in Fig.~\ref{fig:HC-SDE-mu-delta0}(d), and this effect is similar to that of reducing the radius of nanotubes. 
	The similarity is because the former reduces the critical field and the latter reduces the area, both of which make the critical value of flux $\Phi$ decrease. 
	Meanwhile, a smaller $\tilde{\Delta}_0$ and lower temperature reduces the mixture with the states away from the van Hove singularity, which may lead to a larger $|\eta|$ maximum.
	In the experiments, $\Delta_0$ realized by Li-decorated or twisted graphene is from 0.1 meV to 1 meV \cite{ADamascelli-2015-PNAS,CYuan-2018-Nat}, which is much smaller than the 0.1 (eV) used in the previous calculations (limited by the computation resources). Therefore, for chiral carbon nanotubes not as small as one with index (12,6), it is not difficult to observe a complete $\eta(\Phi)$ curve in the $|\Phi|<0.5\Phi_0$ interval.

\section{TMD nanotubes}

    TMD nanotubes have been successfully synthesized and induced to exhibit superconductivity \cite{FQin-2017-NatCom}.
	The 1H-type TMD consists of transition metal atoms sandwiched between two layers of chalcogen atoms, and belongs to the triangular Bravais lattice, the same as graphene.
    The TMDs near the Fermi surface can be described by a three-band model \cite{LGuiBin-2013-PRB} and Eq.~\eqref{eq:tC} is replaced by $t_{\nu \nu'}^\text{TMD}$ given in Eqs. (A3)-(A15).
    
    
    The variation SDE coefficient $\eta$ in  TMD CNTs with respect to temperature $T$ and flux $\Phi$ is shown in Fig.~\ref{fig:TMD-SDE}(a$\sim$c).
    Similar to chiral carbon nanotubes, in a certain temperature range, $\eta$ has a local maxima in weak magnetic fields, changes sign in stronger fields, and reaches a larger $|\eta|$ value.
    
    The TMD has strong spin-orbit coupling, which breaks the inversion symmetry of monolayer TMDs, leading to so called Ising SOC. However, for CNTs, the generation of SDEs does not rely on the breaking of the inversion symmetry in the original 2D structure, as we discussed in Sec.~\ref{sec:CT}. 
    We also verified this by comparing with the SDE of TMD nanotubes without SOC, as shown by the dashed curves in Fig.~\ref{fig:TMD-SDE}(b) and (c). The SDE is quantitatively different due to the change in the shape of the Fermi surface.

    In the calculation, the Fermi surface is considered doped into the valence band of the TMD, as indicated by the dashed line in  Fig.~\ref{fig:TMD-SDE}(d) and (e). When the chemical potential is varied in the neighborhood, the $\eta(\Phi)$ curve changes drastically, up to the order of 10\%, as shown in Fig.~\ref{fig:TMD-SDE}(f). 

	The geometrical index of the chiral nanotubes used is (24,12). According to the experimentally measured lattice constant of about 0.31 nm, the nanotube radius is 1.57 nm, which is roughly the lower limit for experimental fabrication \cite{NYusuke-2023-AdvMat}. 
    For convenience, we use the summation of finite $K$ instead of the integration in the calculation. The results are not affected by this approximation except for extremely low temperatures. 
    The superconducting gap $\tilde{\Delta}_0$ is set to 20 meV for each band, realized by setting different values of the interactions $U$ for different bands. Considering that $\tilde{\Delta}_0$ in TMDs may be an order of magnitude lower than 20 meV, the $\eta(\Phi)$ pattern will shrink horizontally as we showed in the previous section.
    

	\section{Square-lattice materials}	
	
	For comparison, we investigated the SDE of superconducting chiral nanotubes composed of square lattice. The hopping coefficient in Eq.~\eqref{eq:tC} is replaced by $t^\text{Sqr}= -t$ (a number instead of matrix since we assume no internal degrees of freedom other than spin) and the displacement vectors are $\mathbf{d}_1=(1,0)=\hat{\mathbf{a}}_x, \mathbf{d}_2=(0,1)=\hat{\mathbf{a}}_y$.
	We set $\mu=0$ , which is the van Hove singularity at the center of the band.
	
	\begin{figure}
		\centering
		\includegraphics[width=8.5cm]{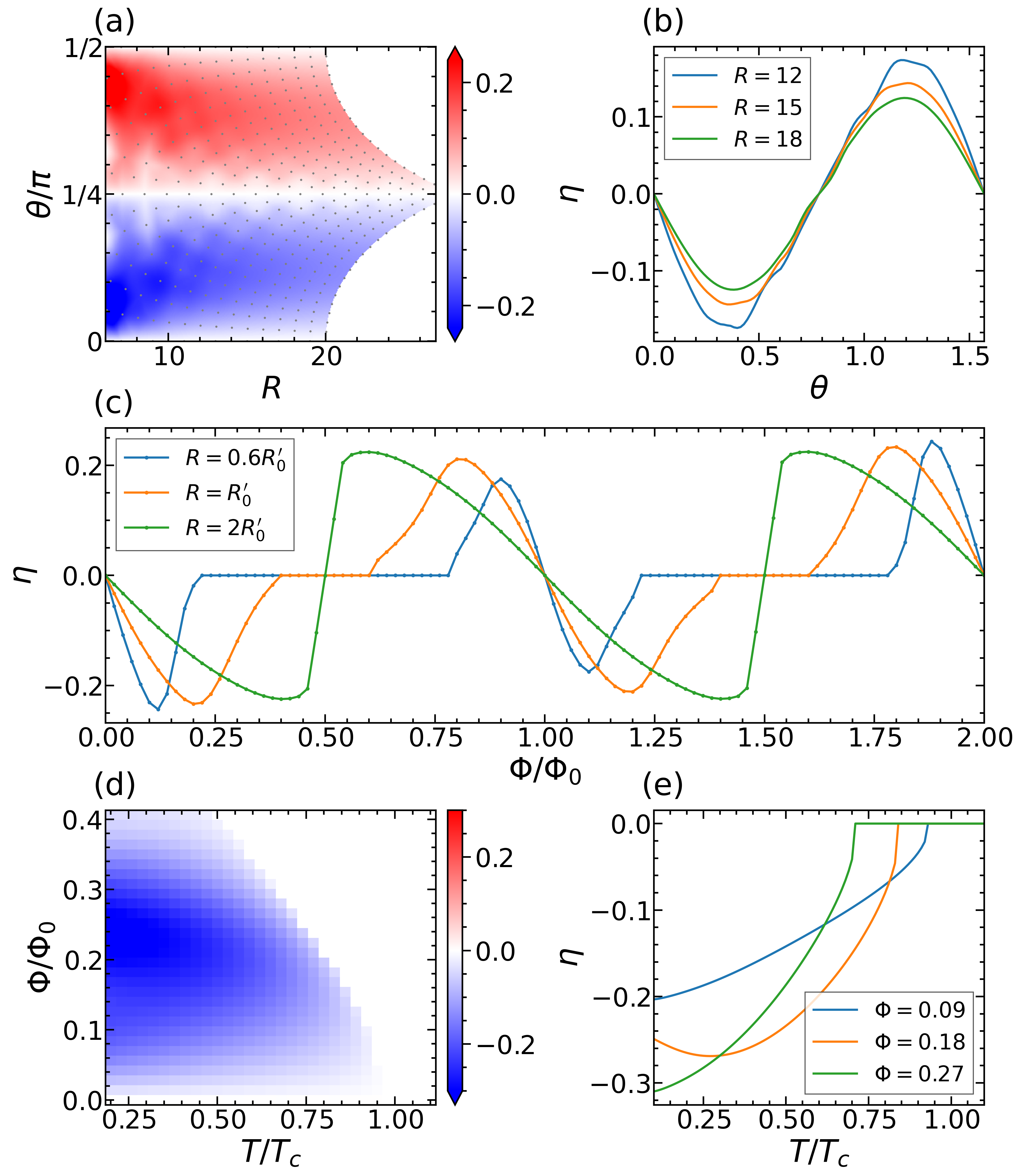}
		\caption{\label{fig:Rect-SDE}SDE in chiral nanotubes made of square lattice. (a) Variation of SDE coefficient $\eta$ with nanotube radius $R$ and chiral angle $\theta$. (b) Variation of $\eta$ with $\theta$ at different radii. (c) Comparison of $\eta(\Phi)$ at different radii, where $R_0'\approx 15.8$ corresponds to the circumferential vector $\mathbf{C} = 15 \hat{\mathbf{a}}_x + 5 \hat{\mathbf{a}}_y$. (d) Variation of $\eta$ with temperature $T$ and magnetic flux $\Phi$. (e) Variation of $\eta$ with $T$ at different $\Phi$. The radius is $R_0'$. Other parameters are $t=1$, $\mu=0$, $\tilde{\Delta}_0=0.2$, $T=0.06$, and $\Phi=0.1\Phi_0$, unless specified.}
	\end{figure}
	
	As shown in Fig.~\ref{fig:Rect-SDE}(a) and (b), the variation of the SDE coefficient $\eta$ with respect to the chiral angle $\theta$ is still approximately sinusoidal. The change of the lattice symmetry (of the unfolded 2D lattice) leads to the a period of $\pi/2$ for the function $\eta(\theta)$.
	The $\eta(\theta)$ curve is smoother than that of chiral carbon nanotubes, which may be related to the simpler $\eta$--$\Phi$ relationship, as shown in Fig.~\ref{fig:Rect-SDE}(c). 
	The variation of the $\eta(\Phi)$ curve with nanotube radius is consistent with the previous calculation, and the $\eta(\Phi)$ pattern may also be truncated due to the flip around $\Phi_0/2$ for large nanotube radius.
	
	The variation of $\eta$ with temperature $T$ and magnetic flux $\Phi$ inside the tube is shown in Fig.~\ref{fig:Rect-SDE}(d). Unlike that of the triangular Bravais lattice, the $|\eta|$ of of the chiral nanotubes with square lattice increases and then decreases with the increase of $\Phi$, and does not change sign in most cases. 
	
	\section{Effects of strains}
	Strain in nanotubes may exist, which shall affect the SDE through changing the band structure. We found that it may enhance or reduce the SDE depending on how it affect the band structure. We investigate discuss this effect in the following, taking carbon CNTs as an example. 
	Since it is relatively easy to apply strain in the axial direction (along the direction of the quasi-one-dimensional tube) after the fabrication, we focus on the strain in this direction.
	Let a strain of proportionality $\varepsilon$ be generated in the axial direction of the nanotube, then the displacement $\mathbf{d}_0$ on the surface of the tube changes to $\mathbf{d}$, \cite{VMPereira-2009-PRB} 
	\begin{align}
		\mathbf{d} =& (1+\boldsymbol{\varepsilon})\cdot\mathbf{d}_0 \\
		\boldsymbol{\varepsilon}=&\varepsilon \begin{pmatrix}
			\cos^2\theta'-\sigma\sin^2\theta' & (1+\sigma)\cos\theta'\sin\theta' \\
			(1+\sigma)\cos\theta'\sin\theta' &\sin^2\theta'-\sigma\cos^2\theta'
		\end{pmatrix}_{x,y} \ .
	\end{align}
	$\sigma$ is the Poisson's ratio, i.e., the proportion of opposite strains generated in the perpendicular direction, and we use the value of graphene $\sigma=0.165$. $\theta'$ is the direction of strain application, {\it i.e.} the direction of the nanotube $\theta'=\theta+\pi/2$.
	The strain causes changes in the lengths of the bonds, which will change the strength of hoppings, thus affecting the Hamiltonian. The hopping strength is \cite{VMPereira-2009-PRB}
	\begin{align}
		t=t_{0}e^{-3.37(d/d_{0}-1)} \ , 
	\end{align}
	where $d_0$ and $d$ are the lattice spacing before and after applying the strain, and $t_0$ and $t$ are the hopping strength before and after applying the strain.
	
	\begin{figure}[!ht]
		\centering
		\includegraphics[width=8.4cm]{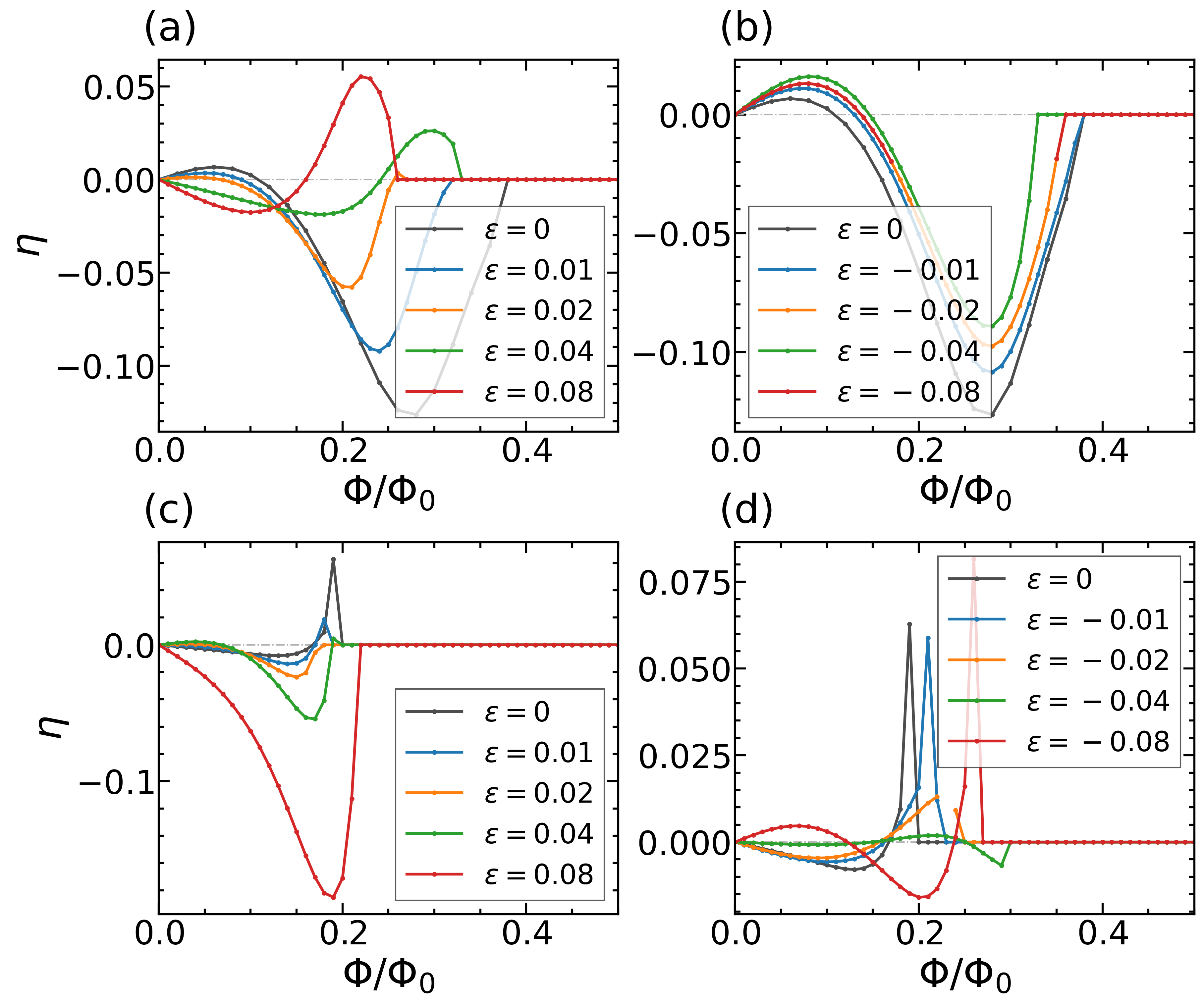}
		\caption{\label{fig:HC-SDE-Strain}Effect of axial strain on SDE of chiral carbon nanotubes. Variation of $\eta(\Phi)$ pattern under (a) different tensile strains and (b) different compressive strains at $\mu=-2.7$ (van Hove singularity). Variation of $\eta(\Phi)$ pattern under (a) different tensile strains and (b) different compressive strains at $\mu=-1.8$. $\varepsilon=0$ is the reference value when no strain is applied. The geometric index is $(m,n)=(12,6)$, other parameters $t_0=2.7$, $\Delta_0=0.1$, $T=0.03$.}
	\end{figure}
	Substituting the parameters after applying strain into the calculation, we get the effect of strain on the SDE $\eta(\Phi)$ pattern of carbon CNTs, as shown in Fig.~\ref{fig:HC-SDE-Strain}.
	Fig.~\ref{fig:HC-SDE-Strain}(a) shows the changes after applying tension, $\eta(\Phi)$ changes more drastically. With the increase of the stretching amplitude, the SDE $|\eta|$ decreases and then flips. The change of $\eta(\Phi)$ applying pressure is more moderate, as shown in Fig.~\ref{fig:HC-SDE-Strain}(b). With increasing pressure, the SDE $\eta$ remains positive when the magnetic field is small and increases, and negative when the magnetic field is large and $|\eta|$ decreases.
	Although the deformation of the Fermi surface produced by strain further breaks the continuous rotational symmetry, it does not increase the maximum value of $|\eta(\Phi)|$. This may be due to that the Fermi surface at the van Hove singularity has the maximum $|\eta(\Phi)|$ amongst the neighboring Fermi levels.
	So we also calculated the effect of tension and pressure on the carbon CNTs at $\mu = -1.8$, as shown in Fig.~\ref{fig:HC-SDE-Strain}(c$\sim$d). The SDE is significantly enhanced when the tension is applied.
	
    \section{Summary and Discussion}
    
    We have shown, using tight-binding models, that both carbon and transition metal dichalcogenide chiral nanotubes (CNTs) under magnetic fields along the tubes exhibit supercurrent diode effect (SDE). Microscopically, this effect results from a discretized energy dispersion made asymmetric by the interplay of the tilted boundary condition (due to the chiral angle) and a transverse momentum shift (due to magnetic flux). 
    The strongest SDE is found when the chemical potential is at the van Hove singularity, where the unfolded band structures have maximum anisotropy, agreeing with previous phenomenological theory. 
    The dependence of the SDE strength on magnetic field and temperature is strong and often show sign flipping at some points.  Square lattice materials are also considered purely for the sake of theoretical comparison and only detailed differences are found ({\it e.g.} the chiral angle periodicity).

    Our study is limited to single-walled nanotubes for simplicity. It is quite obvious that multi-walled CNTs should lead to the same effect because the key is the asymmetry of the energy dispersion, which can also be present in multi-walled CNTs as long as the multiple walls are not align in such a way that the chiral geometries in each of them cancel with each other.

    Carbon CNTs are either non-superconducting or the $T_c$ is too low. In this case, our results apply to those carbon CNTs in proximity to superconductors. For TMD CNTs, they are superconducting with $T_c$ being several Kelvin, thus the SDE should be observable. In this case, the SDE due to chiral the structure should not be confused with those from Ising spin-orbit coupling. 
    The fact that the SDE is due to the chiral structure can be seen in its dependence on the chiral angle, where the SDE vanishes at $\theta=n\pi/6$.  
    
    
    \section*{Acknowledgement}
    We used the Kwant package for building Hamiltonian matrix \cite{Groth-2014-NJP}. 
    JJH is supported by the Innovation Program for Quantum Science and Technology (Grant No. 2021ZD0302800) and the National Natural Science Foundation of China (Grant No. 12204451). 
    
    \appendix
    
    \section{The Hopping terms of TMD} \label{Appendix1}
    The reference \cite{LGuiBin-2013-PRB} and its erratum have given the three-band tight-binding Hamiltonian for monolayers of 1H-type TMD in $\mathbf{k}$ space.
    In this section, we write the hopping terms in real space.
    
    With basis $(d_{z^2}, d_{xy}, d_{x^2-y^2})$, 
    the generators of $D_{3h}$ group are the three-fold rotation
    \begin{align}
    	\hat{C}_3 &= \begin{pmatrix}
    		1 & 0 & 0 \\
    		0 & -\frac{1}{2} & -\frac{\sqrt{3}}{2} \\
    		0 & \frac{\sqrt{3}}{2} & -\frac{1}{2}
    	\end{pmatrix} 
   \end{align}
   and the mirror
    \begin{align}
    	\hat{\sigma}_v &= \begin{pmatrix}
    		1 & 0 & 0 \\
    		0 & \frac{1}{2} & \frac{\sqrt{3}}{2} \\
    		0 & \frac{\sqrt{3}}{2} & -\frac{1}{2}
    	\end{pmatrix} \ .
    \end{align}
    The onsite submatrix is $t_{\nu\nu'}^\text{TMD}(\mathbf{0})= \text{diag}\{ \epsilon_1, \epsilon_2, \epsilon_2 \}$.
    There are six nearest-neighbor hoppings with the displacements $\mathbf{d}_1 =(1,0)$, $\mathbf{d}_2 =(1/2,\sqrt{3}/2)$, $\mathbf{d}_3 =(-1/2,\sqrt{3}/2)$, $\mathbf{d}_4 =-\mathbf{d}_1$, $\mathbf{d}_5 =-\mathbf{d}_2$, and $\mathbf{d}_6 =-\mathbf{d}_3$.
    And the nearest-neighbor hopping submatrices in Hamiltonian are
    \begin{widetext}		
    	\begin{align}
    		t_{\nu\nu'}^\text{TMD}(\mathbf{d}_1) 
    		=& \begin{pmatrix}
    			t_{0} & -t_{1} & t_{2} \\
    			t_{1} & t_{11} & -t_{12} \\
    			t_{2} & t_{12} & t_{22} \\
    		\end{pmatrix}_{\nu\nu'} 
    		\\
    		\begin{split}
    		t_{\nu\nu'}^\text{TMD}(\mathbf{d}_4) =& (\hat{C}_3 \hat{\sigma}_v) t_{\nu\nu'}^\text{TMD}(\mathbf{d}_1) (\hat{C}_3 \hat{\sigma}_v)^\dagger \\ 
    		=& [t_{\nu\nu'}^\text{TMD}(\mathbf{d}_1)]^\dagger
    		\end{split}\\
    		\begin{split}
    		t_{\nu\nu'}^\text{TMD}(\mathbf{d}_2) =& \hat{\sigma}_v t_{\nu\nu'}^\text{TMD}(\mathbf{d}_1) \hat{\sigma}_v^\dagger \\
    		=& \begin{pmatrix}
    			t_{0} & \frac{1}{2} (-t_{1}+\sqrt{3} t_{2}) & \frac{1}{2} (-\sqrt{3} t_{1}-t_{2}) \\
    			\frac{1}{2} (t_{1}+\sqrt{3} t_{2}) & \frac{1}{4} (t_{11}+3 t_{22}) & \frac{\sqrt{3}}{4} (t_{11}-t_{22})+t_{12} \\
    			\frac{1}{2} (\sqrt{3} t_{1}-t_{2}) & \frac{\sqrt{3}}{4} (t_{11}-t_{22})-t_{12} & \frac{1}{4} (3 t_{11}+t_{22}) \\
    		\end{pmatrix}_{\nu\nu'}
		    \end{split} \\
		    \begin{split}
    		t_{\nu\nu'}^\text{TMD}(\mathbf{d}_5) =& (\hat{C}_3 \hat{C}_3) t_{\nu\nu'}^\text{TMD}(\mathbf{d}_1) (\hat{C}_3 \hat{C}_3)^\dagger 
    		= [t_{\nu\nu'}^\text{TMD}(\mathbf{d}_2)]^\dagger
    		\end{split} \\
    		\begin{split}
    		t_{\nu\nu'}^\text{TMD}(\mathbf{d}_3) =& \hat{C}_3 t_{\nu\nu'}^\text{TMD}(\mathbf{d}_1) \hat{C}_3^\dagger \\
    		=& \begin{pmatrix}
    			t_{0} & \frac{1}{2} (t_{1}-\sqrt{3} t_{2}) & \frac{1}{2} (-\sqrt{3} t_{1}-t_{2}) \\
    			\frac{1}{2} (-t_{1}-\sqrt{3} t_{2}) & \frac{1}{4} (t_{11}+3 t_{22}) & \frac{\sqrt{3}}{4} (-t_{11}+t_{22})-t_{12} \\
    			\frac{1}{2} (\sqrt{3} t_{1}-t_{2}) & \frac{\sqrt{3}}{4} (-t_{11}+t_{22})+t_{12} & \frac{1}{4} (3 t_{11}+t_{22}) \\
    		\end{pmatrix}_{\nu\nu'}
		    \end{split} \\
		    \begin{split}
    		t_{\nu\nu'}^\text{TMD}(\mathbf{d}_6) =& (\hat{\sigma}_v \hat{C}_3) t_{\nu\nu'}^\text{TMD}(\mathbf{d}_1) (\hat{\sigma}_v \hat{C}_3)^\dagger 
    		= [t_{\nu\nu'}^\text{TMD}(\mathbf{d}_3)]^\dagger
    		\end{split}.
    	\end{align}
    \end{widetext}
    There are six next-nearest-neighbor hoppings with the displacements $\pm(\mathbf{d}_1+\mathbf{d}_2)$, $\pm(\mathbf{d}_3+\mathbf{d}_4)$, and $\pm(\mathbf{d}_5+\mathbf{d}_6)$.
    The next-nearest-neighbor hopping submatices in Hamiltonian are
    	\begin{align}
    		t_{\nu\nu'}^\text{TMD}(\mathbf{d}_1+\mathbf{d}_2) 
    		=& \begin{pmatrix}
    			r_{0} & -r_{1} & -\frac{1}{\sqrt{3}}r_{1} \\
    			-r_{2} & r_{11} & -r_{12} \\
    			-\frac{1}{\sqrt{3}}r_{2} & -r_{12} & r_{11}+\frac{2}{\sqrt{3}}r_{12}
    		\end{pmatrix}_{\nu\nu'} \notag\\
    		=& \hat{\sigma}_v t_{\nu\nu'}^\text{TMD}(\mathbf{d}_1+\mathbf{d}_2) \hat{\sigma}_v^\dagger
    		 \\
    		t_{\nu\nu'}^\text{TMD}(-\mathbf{d}_1-\mathbf{d}_2)
    		=& [t_{\nu\nu'}^\text{TMD}(\mathbf{d}_1+\mathbf{d}_2)]^\dagger \\
    		t_{\nu\nu'}^\text{TMD}(\mathbf{d}_3+\mathbf{d}_4) =& \hat{C}_3  t_{\nu\nu'}^\text{TMD}(\mathbf{d}_1+\mathbf{d}_2) \hat{C}_3^\dagger \notag \\
    		=& \begin{pmatrix}
    			r_{0} & r_{1} & -\frac{1}{\sqrt{3}}r_{1} \\
    			r_{2} & r_{11} & r_{12} \\
    			-\frac{1}{\sqrt{3}}r_{2} & r_{12} & r_{11}+\frac{2}{\sqrt{3}}r_{12} \\
    		\end{pmatrix}_{\nu\nu'} \notag\\
    		=& (\hat{C}_3 \hat{\sigma}_v) t_{\nu\nu'}^\text{TMD}(\mathbf{d}_1+\mathbf{d}_2) (\hat{C}_3 \hat{\sigma}_v)^\dagger
 \\
    		t_{\nu\nu'}^\text{TMD}(-\mathbf{d}_3-\mathbf{d}_4) 
    		=& [t_{\nu\nu'}^\text{TMD}(\mathbf{d}_3+\mathbf{d}_4)]^\dagger  
    		\\
    		t_{\nu\nu'}^\text{TMD}(\mathbf{d}_5+\mathbf{d}_6) =& (\hat{C}_3 \hat{C}_3) t_{\nu\nu'}^\text{TMD}(\mathbf{d}_1+\mathbf{d}_2) (\hat{C}_3 \hat{C}_3)^\dagger \notag \\
    		=& \begin{pmatrix}
    			r_{0} & 0 & \frac{2}{\sqrt{3}}r_{1} \\
    			0 & r_{11}+\sqrt{3} r_{12} & 0 \\
    			\frac{2}{\sqrt{3}}r_{2} & 0 & r_{11}-\frac{1}{\sqrt{3}}r_{12} \\
    		\end{pmatrix}_{\nu\nu'} \notag\\
    		=& (\hat{\sigma}_v \hat{C}_3) t_{\nu\nu'}^\text{TMD}(\mathbf{d}_1+\mathbf{d}_2) (\hat{\sigma}_v \hat{C}_3)^\dagger
    		\\
    		t_{\nu\nu'}^\text{TMD}(-\mathbf{d}_5-\mathbf{d}_6)
    		=& [t_{\nu\nu'}^\text{TMD}(\mathbf{d}_5+\mathbf{d}_6)]^\dagger \ .
    	\end{align}
    The third-nearest-neighbor hoppings is similar to the nearest-neighbor hoppings, and can be obtained by $\mathbf{d}_i \to 2\mathbf{d}_i$, and $t_\nu \to u_\nu$.
    
    By Fourier transform $c_{\mathbf{r}}^\dagger = \sum_{\mathbf{r}} e^{-i\mathbf{k}\cdot\mathbf{r}} c_{\mathbf{k}}^\dagger$, we can restore the $\mathbf{k}$-space Hamiltonian $H^\text{TNN}(\mathbf{k})$ in reference \cite{LGuiBin-2013-PRB}, with the same notation for the coefficients.
	
	\section{Fix-Point Iteration \& Anderson Acceleration} \label{Appendix2}
	Before calculating the current in a superconducting nanotube, we need to determine the superconducting gap $\Delta$ for a specific Cooper pair momentum $q$. 
	For $\Delta_\nu$ at each position and for each orbit, we can substitute the trial solution into the Hamiltonian $H$, and through the gap equation
	\begin{subequations}
		\begin{align}
			\Delta_\nu
			=& -U \langle c_{\nu \downarrow} c_{\nu \uparrow} \rangle \\
			=& -U \sum_\alpha \langle \psi_\alpha | \frac{\partial H}{\partial \Delta_\nu^*} | \psi_\alpha \rangle f_D(E_\alpha) \ ,
		\end{align}
	\end{subequations}
	compute the new value and use it as the next trial solution, iterating until it converges to the fixed point $\Delta_\nu$. 
	Here, $U>0$ is the attractive interaction between electrons, $E_\alpha$ and $| \psi_\alpha \rangle$ are the eigenenergies and eigenstates of $H$, and $f_D(E_\alpha)=[\exp(E/k_BT)+1]^{-1}$ is the Fermi--Dirac distribution.
	The gap equation can be derived by the condition that the free energy $F=-k_BT\sum_\alpha \ln(1+\exp(E_\alpha/k_BT))+\sum_\nu |\Delta_\nu|^2/U$ takes a local minimum at $\Delta_\nu$,
	\begin{align} \label{eq:Min-FreeE}
		0 = \frac{\partial F}{\partial \Delta_\nu^*}
		= \langle c_{\nu \downarrow} c_{\nu \uparrow} \rangle +\frac{\Delta_\nu}{U} \ .
	\end{align}
	
	Analogous to Eq.~\eqref{eq:Min-FreeE}, in the (k+1)th iteration
	\begin{subequations}
		\begin{align}
			\Delta_\nu^{(k+1)} 
			=& -U \langle c_{\nu\downarrow} c_{\nu\uparrow} \rangle^{(k)} \\
			=& \Delta_\nu^{(k)} -U \left( \frac{\partial F}{\partial \Delta_\nu^*} \right)^{(k)} \ .
		\end{align}
	\end{subequations}
	Therefore, the algorithm for the fixed point iteration is equivalent to a gradient descent algorithm with $F$ as the objective function and a fixed step size $U/2$ in the complex plane of $\Delta_\nu$.
	Using Anderson's acceleration algorithm we can speed up the stationary point iteration process, which is equivalent to correcting the step size based on historical calculations.
	
	Denote $u^{(k)}$ as the vector composed of $\{\Delta_\nu\}$ in the (k)th iteration, and $f$ as the iterative computation process, the Anderson's acceleration algorithm steps as follows.
	Starting from the initial point $u^{(0)}$, we iterate to get $f(u^{(0)})$, and the difference is $d^{(0)} = f(u^{(0)}) - u^{(0)}$.
	In the first iteration, we use $u^{(1)}=f(u^{(0)})$, compute $f(u^{(1)})$, and get the difference $d^{(1)}$.
	In the (k+1)th iteration ($k=1,\,2,\,\dots$), we take the last $(m+1)$ results ($m= \min\{k,m_0\}$, $m_0$ is usually less than 5), and consider the weighted average of the differences
	\begin{align} \label{eq:diff_avg}
		\bar{d} = \sum_{j=0}^{m} \alpha_j d^{(k-j)} \ .
	\end{align}
	The weight $\sum_{j=0}^{m} \alpha_j =1$, $\alpha_j \in \mathbb{R}$ can be positive or negative.
	Minimize $|\bar{d}|^2$ to get $\{ \alpha_j \}$, and let $u^{(k+1)}= \sum_{j=0}^{m} \alpha_j f(u^{(k-j)})$.
	Then, we substitute $u^{(k+1)}$, $f(u^{(k+1)})$, $d^{(k+1)}$ into the next iteration until convergence.
	
	The $\{ \alpha_j \}$ is computed as follows. 
	Substituting the constraints of $\{ \alpha_j \}$ into Eq.~\eqref{eq:diff_avg}, the problem is transformed to minimize
	\begin{align}
		|\bar{d}|^2 = |d^{(k)} - \sum_{j=1}^{m} \alpha_j (d^{(k)} - d^{(k-j)}) |^2 \ ,
	\end{align}
	without constraints.
	Convert to matrix form: $X=(\alpha_1 \cdots \alpha_m)^T$; $Y=d^{(k)}=(d^{(k)}_1 \cdots d^{(k)}_\sigma)^T$, $\sigma$ is the dimension of $u^{(k)}$ and $d^{(k)}$; $A=((d^{(k)} - d^{(k-1)}) \cdots (d^{(k)} - d^{(k-m)}))$, a matrix of dimension $\sigma \times m$.
	\begin{subequations}
		\begin{align}
			|\bar{d}|^2 =& (Y-AX)^\dagger(Y-AX) \\
			=& Y^\dagger Y +X^T A^\dagger AX -Y^\dagger AX -X^TA^\dagger Y \ ,
		\end{align}
	\end{subequations}
	Derive $|\bar{d}|^2$ with respect to $X$ at extremum,
	\begin{subequations}
		\begin{align}
			0 =& \begin{pmatrix}
				\frac{\partial}{\partial \alpha_1} & \cdots & \frac{\partial}{\partial \alpha_m}
			\end{pmatrix} |\bar{d}|^2 \\
			=& [A^\dagger AX +(X^TA^\dagger A)^T] -(Y^\dagger A)^T -A^\dagger Y \\
			=& (A^\dagger A +A^T A^*)X -A^T Y^* -A^\dagger Y \ ,
		\end{align}
	\end{subequations}
	we obtain
	\begin{align}
		X= ( \text{Re}[A^\dagger A] )^{-1} \text{Re}[A^\dagger Y] \ .
	\end{align}
	
	\bibliography{ref}
	
\end{document}